\documentclass[12pt]{article}
\usepackage{epsfig,a4}
\usepackage{color}
\begin{document}
\title{Color symmetrical superconductivity in a
schematic nuclear quark model.}
\author{
Henrik Bohr\\
{\it \small Department of Physics, B.307, Danish Technical
University,}\\{ \it \small DK-2800 Lyngby, Denmark}\\
Jo\~ao da Provid\^encia\\
{\it \small Departamento de F\'\i sica, Universidade de Coimbra,}\\
{\it \small  P-3004-516 Coimbra, Portugal}  }


\maketitle
\def\e{\rm e}\def\d{{\rm d}}
\def\oln{\overline{n}}
\def\ve{\varepsilon}
\def\bfp{{\bf p}}
\def\bfx{{\bf x}}
\def\bfk{{\bf k}}
\def\bfq{{\bf q}}
\def\wtn{\widetilde{n}}
\def\wtt{\widetilde\theta}
\def\ol{\overline}
\def\a{\alpha}
\def\b{\beta}
\abstract{In this paper, the phenomenon of color superconductivity
is analyzed on the basis of  a novel BCS-type formalism and in the
context  of a schematic QCD inspired quark model, having in mind the
description of color symmetrical superconducting states. The
QCD-like model has color $SU(3)$ symmetry and is constructed in
terms of the generators of the $su(4)$ algebra, becoming exactly
solvable. The physical properties of the BCS vacuum (average numbers
of quarks of different colors) remain unchanged under an arbitrary
color rotation. In the usual approach to color superconductivity,
the pairing correlations affect only the quasi-particle states of
two colors, the single particle states of the third color remaining
unaffected by the pairing correlations. In the theory of color
symmetrical superconductivity here proposed, the pairing
correlations affect symmetrically the quasi-particle states of the
three colors so that vanishing net color-charge is automatically
insured. It is found that the ground-state energy of the color
symmetrical sector of this exactly solvable model is well
approximated by the average energy of the color symmetrical
superconducting state proposed here.}

\section{Introduction}
At present, it is generally accepted that QCD matter at high
densities exhibits color superconductivity induced by the familiar
phenomenon of Cooper instability \cite{alford}. For a recent review,
see \cite{alford1}.  That color superconductivity breaks color
$SU(3)$ symmetry down to the $SU(2)$ level is a familiar statement,
which, however, may require further specification. Quarks are free
in the deconfined phase, but the deconfined phase itself is believed
to be a color singlet. For instance, in Refs.
\cite{gerhold,dietrich} it is argued that in QCD the superconducting
phase is automatically color symmetrical. \footnote{The exressions
``color singlet", ``color symmetrical" and ``color neutral", are
here used as synonymous, and refer to states belonging to one
dimensional representation of color $SU(3)$.} Our aim is to develop
a version of the BCS theory which is appropriate to describe a color
singlet superconducting phase.  The physical properties of the BCS
vacuum which is here constructed (average numbers of quarks of
different colors) remain unchanged under an arbitrary color
rotation.  The problem of reconciling color superconductivity with
color symmetry has been addressed by some authors \cite{amore,iida}.
In \cite{amore}, the authors resort to a rather involved projection
technique to extract color symmetrical states out of BCS sates which
violate color symmetry. In \cite{iida}, color symmetry is imposed by
requiring that the average or expectation value of some of the eight
color generators vanish, that is, color neutrality is implemented in
the average.  The approach followed in \cite{amore} suffers from an
obvious deficiency which is known to be associated with the
Peierls-Yoccoz projection method, namely, the reliability of the
result obtained depends strongly on the choice of the meanfield
state to which the method is applied (see \cite{ring}, pp. 460-462).
That meanfield state must to be properly chosen, that is, it has to
be chosen in such a way that you may reasonably expect the
components with the desired symmetry to have a dominant
contribution. Otherwise you end up by extracting a component which,
of course, has the desired symmetry but which most probably has
nothing to do with the state you are looking for. For instance,
suppose you are studying a deformed nucleus and wish to describe the
lowest energy state with angular momentum $J$  by applying the
Peierls-Yoccoz projection method to a meanfield state vector. You
may easily extract a state with the desired angular momentum which,
however, most probably will lead to the wrong moment of inertia. The
projection method is reliable only if the mean field state has been
constructed according to very precise rules, following a procedure
analogous to the one we will describe in the sequel. Thus, the
following development is of interest even if you have in mind using
the projection method, as in \cite{amore}.

A BCS state $|\Phi\rangle$ describes a physical state with zero net
color charge if $N_1=N_2=N_3$, where $N_i$ denotes the average
number of quarks of color $i$. This means that
\begin{eqnarray}\langle\Phi|S_{11}|\Phi\rangle=\langle\Phi|S_{22}|\Phi\rangle
=\langle\Phi|S_{33}|\Phi\rangle.\label{cneutral}
\end{eqnarray} Here, $S_{kl}$ denote the color $U(3)$ generator
which will be defined in the next section. However, we will show
that the requirement (\ref{cneutral}), which is implemented in
\cite{iida}, is not sufficient to insure that $|\Phi\rangle$
describes a color singlet, which is the condition for physical
acceptability. A stronger condition must then be imposed. Indeed,
the $SU(3)$ symmetry, being a gauge symmetry, cannot be broken,
according to the discussion in \cite{gerhold,dietrich}. Therefore,
color rotated BCS states must be equivalent in the sense of the
physics they describe. Let $U_c$ denote an arbitrary color rotation,
i.e., $U_c=\exp \sum_{k,l=1}^3ix_{kl}S_{kl}$, $x_{kl}=x_{lk}^*$. The
BCS state $|\Phi\rangle$ must be equivalent to the state
$U_c|\Phi\rangle$, for any $U_c$, as far as expectation values of
physical observables are concerned. Thus, the condition
(\ref{cneutral}) must be replaced by
\begin{eqnarray*}\langle\Phi|U_c^\dagger
S_{11}U_c|\Phi\rangle=\langle\Phi|U_c^\dagger
S_{22}U_c|\Phi\rangle=\langle\Phi|U_c^\dagger
S_{33}U_c|\Phi\rangle,\end{eqnarray*} for an arbitrary $U_c$, and
this implies
\begin{eqnarray}\langle\Phi|S_{11}|\Phi\rangle=\langle\Phi|S_{22}|\Phi\rangle
=\langle\Phi|S_{33}|\Phi\rangle,\quad\langle\Phi|S_{kl}|\Phi\rangle=0,{\rm~~~
for~~~} k\neq l.\label{csinglet}\end{eqnarray} This is the condition
the BCS state $|\Phi\rangle$ must necessarily satisfy in order to be
physically meaningful. If only the condition (\ref{cneutral}) is
implemented, as in \cite{iida}, and not the condition
(\ref{csinglet}), the BCS state $|\Phi\rangle$ is, in general, not
equivalent to the state $U_c|\Phi\rangle$, so that it describes a
state belonging to a representation of $SU(3)$ other than the
singlet one, which is physically unacceptable. In \cite{bohr}, it
has already been suggested that the condition (\ref{csinglet})
should be satisfied if we wish that the BCS state $|\Phi\rangle$
describes a color singlet state. Here, we show how this condition
may be easily implemented.

We approach this problem by constructing the BCS state through an
appropriate generalization of the Bogoliubov transformation that
treats all colors on the same footing. For simplicity, the present
derivation is performed in the framework of a standard nuclear quark
model exemplified by the Bonn model which was proposed by H.R. Petry
{\it et al.} in 1985 \cite{petry}, but the scope of the formalism
here obtained is quite general. This model is defined through a
specific color dependent pairing interaction which is expressed in
terms of certain generators of $SU(4)$ and is invariant under
$SU(3)$. Due to this important symmetry, that model is usually
regarded as a model for the formation of color symmetrical triplets.
In ref. \cite{bohr}, exact solutions have been presented to the
equations of the model, it being shown that, in general, its
groundstate is not color symmetrical, although it admits an
important color symmetrical sector. Using a representation of
$su(4)$ of the Schwinger type which is formulated in terms of
appropriate boson operators and is due to Yamamura {\it at al.}
\cite{yamamura}, it has been possible to characterize the referred
color symmetrical sector \cite{c.providencia}. In ref. \cite{bohr},
it has been shown that BCS states describe adequately the
groundstate of the model. In this note we focus on the description
of the color symmetrical sector of the model.

\section{A schematic $su(3)$  pairing  model}

The quark model proposed by H.R. Petry {\it et al.} \cite{petry} is
defined by the Hamiltonian
\begin{equation}\label{H}H=G\sum_{j=1}^3A^\dagger_jA_j,\end{equation} where
\begin{equation}\label{2}A^\dagger_1=\sum_{m>0}( c^\dagger_{2m}c^\dagger_{3\ol
m}+c^\dagger_{2\ol m}c^\dagger_{3 m}) , \quad \ol{\ol
m}=m,\end{equation} $G<0$ is the coupling constant and the
expressions for $A^\dagger_{2},A^\dagger_{3},$ are obtained by
circular permutation of the indices $1,2,3$. In eq. (\ref{2}), $c_{i
m}^\dagger$ are quark creation operators and the indices $i$ and $m$
denote, respectively,  the color and the remaining single particle
quantum numbers. By $\ol m$ we mean the state obtained from $m$ by
time reversal. We remark, in passing, that this interaction is of
the same type  as the one used to describe color superconducting
quark matter \cite{alford}. The present model is an effective QCD
model in the limit of asymptotic freedom of vanishing nuclear
forces. In a future study we shall include the barions as a
three-body force.

Color superconductivity has been applied in \cite{bohr} to the
description of the groundstate of the Bonn model, which in general
is not color symmetric. Indeed, although $H$ has $SU(3)$ symmetry,
its eigenstates are not necessarily invariant under color $SU(3)$
rotations, that is, they are not necessarily color singlets. The
study of the color symmetrical sector is particularly interesting.
The generators of color $U(3)$ read
$$S_{kl}=\sum_m c^\dagger_{km}c_{lm}=\sum_{m>0} (c^\dagger_{km}c_{lm}+c^\dagger_{k\ol m}c_{l\ol m}).
$$
A state $|\Phi\rangle$ is a color singlet if it satisfies the
following condition
\begin{equation}
S_{kl}|\Phi\rangle=0,\quad k\neq l,\quad S_{kk}|\Phi\rangle=\lambda
|\Phi\rangle,\quad k=1,2,3.\label{CNC}
\end{equation}
A BCS state can not satisfy (\ref{CNC}), so that, strictly speaking,
it can not be a color singlet, but it will describe a color singlet
if, and only if, it satisfies (\ref{csinglet}), which amounts to
satisfying (\ref{CNC}) in the average. It has been remarked in
\cite{bohr} that the description of the color symmetrical sector of
the model requires a modified BCS state, defined as the vacuum of
the quasi-particles of an appropriate generalized Bogoliubov
transformation.

 \section{ Color symmetric BCS state}
The most general case  
BCS state which, in the average, is color neutral,  reads
$$|\Phi\rangle=\exp\sum_{j=0}^3\left(K\sum_{0<m\leq\Omega'} A^\dagger_{jm}
+\tilde K\sum_{\Omega'<m\leq\Omega} A_{jm}\right)|
0_{\Omega'}\rangle,$$where
$$ | 0_{\Omega'}\rangle=\left(\prod_{j=1}^3\prod_{\Omega'
< m\leq\Omega}c^\dagger_{jm}c^\dagger_{j\ol m}\right)|0\rangle,$$
and
$$ A^\dagger_{1m}=(c^\dagger_{2m} c^\dagger_{3\ol m}+c^\dagger_{2\ol m}
c^\dagger_{3 m}).
$$
This state is color neutral in the sense of vanishing net color
charge. The expressions for $A^\dagger_{2m},A^\dagger_{3m},$ are
obtained by circular permutation of the indices $1,2,3$. The
parameters $K,~\tilde K$ are real. We denote by $2\Omega$ the level
degeneracy for a fixed color, that is, the totality of eigenstates
pertaining to all quantum numbers beyond color. If $\Omega'=\Omega,$
the quark number $N$ satisfies $0\leq N\leq4\Omega$. If $\Omega'=0,$
the quark number $N$ satisfies $4\Omega\leq N\leq6\Omega$. The state
vector $|\Phi\rangle$ has obviously zero net color charge, but it is
not color symmetrical. This is because
$[S_{12},(A^\dagger_1+A^\dagger_2+A^\dagger_3)]=-A^\dagger_2\neq 0.
$ However, $K,~\tilde K$ may be chosen so that $|\Phi\rangle$ is
color symmetrical in the average, that is, so that (\ref{cneutral})
is satisfied. We observe that
\begin{eqnarray}&&c_{1m}|\Phi\rangle=K(c^\dagger_{2\ol
m}-c^\dagger_{3\ol m})|\Phi\rangle, \quad c_{1\ol
m}|\Phi\rangle=K(c^\dagger_{2 m}-c^\dagger_{3 m})|\Phi\rangle,\quad
0<m\leq\Omega',\nonumber\\&& c^\dagger_{1m}|\Phi\rangle=-\tilde
K(c_{2\ol m}-c_{3\ol m})|\Phi\rangle, \quad c^\dagger_{1\ol
m}|\Phi\rangle=-\tilde K(c_{2 m}-c_{3 m})|\Phi\rangle,\quad
\Omega'<m\leq\Omega.\nonumber\\&&\label{3}
\end{eqnarray} These relations are crucial. They are straightforward
consequences of the commutation relations
\begin{eqnarray*}&&\left[c_{1p}, \left(K\sum_{0<m\leq\Omega'}
A^\dagger_{jm} +\tilde K\sum_{\Omega'<m\leq\Omega}
A_{jm}\right)\right]=K(c^\dagger_{2\ol p}-c^\dagger_{3\ol p}),\quad
0<p\leq\Omega',\\&& \left[c_{1p}^\dagger,
\left(K\sum_{0<m\leq\Omega'} A_{jm}^\dagger +\tilde
K\sum_{\Omega'<m\leq\Omega} A_{jm}\right)\right]=-\tilde K(c_{2\ol
p}-c_{3\ol p}),\quad \Omega'<p\leq\Omega.
\end{eqnarray*}
From (\ref{3}) it follows that the BCS vacuum $|\Phi\rangle$ is
annihilated by the operators
\begin{eqnarray}&&d_{1m}=c_{1m}-K(c^\dagger_{2\ol m}-c^\dagger_{3\ol m}),
\quad d_{1\ol m}=c_{1\ol m}-K(c^\dagger_{2 m}-c^\dagger_{3 m}),\quad
0<m\leq\Omega',\nonumber\\&& d_{1m}=c^\dagger_{1m}+\tilde K(c_{2\ol
m}-c_{3\ol m}), \quad d_{1\ol m}= c^\dagger_{1\ol m}+\tilde K(c_{2
m}-c_{3 m}),\quad \Omega'<m\leq\Omega.\nonumber\\&&\label{4}
\end{eqnarray} The expressions for $d_{2m},~d_{3m},~d_{2\ol
m},~d_{3\ol m},$ are obtained by circular permutation of the indices
1,2,3. These operators characterize the so-called Bogoliubov
quasi-particles. The transformation in eq. (\ref{4}) is not
canonical, since $\{d_{im},d_{jm}^\dagger\}\neq\delta_{ij}$, but the
corresponding canonical transformation, which is not needed for the
present purpose, may be easily obtained.  \footnote{A generalized
Bogoliubov transformation has been proposed in eq. (25) of
\cite{bohr}. However, the associated BCS state does not satisfy
(\ref{csinglet}). The transformation (\ref{4}) replaces eq. (25) of
\cite{bohr}.}

We introduce the notation $\langle
W\rangle=\langle\Phi|W|\Phi\rangle/\langle\Phi|\Phi\rangle$. We
observe that the contractions $\langle c_{im}^\dagger
c_{jm}\rangle,~i\neq j$, are independent of $i,~j$. Similarly the
contractions $\langle c_{jm}^\dagger c_{jm}\rangle,$ are independent
of $j$.

We easily find
$$\langle c_{im}^\dagger c_{jm}\rangle=\langle c_{i\ol m}^\dagger
c_{j\ol m}\rangle=-{K^2\over1+3 K^2},~i\neq
j,\quad \langle c_{jm}^\dagger c_{jm}\rangle=\langle c_{j\ol
m}^\dagger c_{j\ol m}\rangle={2K^2\over1+3K^2},$$ if $
0<m\leq\Omega'. $

On the other hand,
$$\langle c_{im}^\dagger c_{jm}\rangle=\langle c_{i\ol m}^\dagger c_{j\ol m}\rangle={\tilde K^2\over1+3 \tilde K^2},~i\neq
j,\quad \langle c_{jm}^\dagger c_{jm}\rangle=\langle c_{j\ol
m}^\dagger c_{j\ol m}\rangle=1-{2\tilde K^2\over1+3\tilde K^2},$$ if
$ \Omega'<m\leq\Omega. $

These results are obtained as follows. For $0<m\leq\Omega'$, we have
$$X:=\langle c^\dagger_{1m}c_{2m}\rangle=-K^2-K^2(-\langle c^\dagger_{3\ol m} c_{3\ol m}\rangle-\langle c^\dagger_{2\ol m} c_{1\ol m}\rangle
+\langle c^\dagger_{3\ol m} c_{1\ol m}\rangle+\langle
c^\dagger_{2\ol m} c_{3\ol m}\rangle)
$$
$$N:=\langle c^\dagger_{1m}c_{1m}\rangle=2K^2-K^2(\langle c^\dagger_{3\ol m} c_{3\ol m}\rangle+\langle c^\dagger_{2\ol m} c_{2\ol m}\rangle
-\langle c^\dagger_{3\ol m} c_{2\ol m}\rangle-\langle
c^\dagger_{2\ol m} c_{3\ol m}\rangle),
$$ implying
$$X=-K^2+K^2(N-X),\quad N=2K^2-2K^2(N-X),
$$ which leads to $X=-K^2/(1+3 K^2),\;N=2K^2/(1+3 K^2). $ The corresponding
expressions for $\Omega'<m\leq\Omega$ are similarly obtained. We
find
$$\langle S_{ij}\rangle=-2\Omega'{K^2\over1+3K^2}+2(\Omega-\Omega'){\tilde K^2\over1+3\tilde
K^2},\quad i\neq j.
$$
By conveniently choosing $K,\,\tilde K$, we may insure that the
condition for the BCS vacuum $|\Phi\rangle$ to be a  color singlet,
namely
\begin{eqnarray}\langle\Phi| S_{ij}|\Phi\rangle=0,\quad{\rm for}\quad i\neq
j,\label{csym}
\end{eqnarray} is satisfied. The state $|\Phi\rangle$ satisfies
automatically (\ref{cneutral}). However, (\ref{cneutral}) remains
valid when we replace $|\Phi\rangle$ by $U_c|\Phi\rangle$, for an
arbitrary color rotation $U_c$, only if (\ref{csym}) is further
imposed.

Next we compute the contractions $ \langle c_{2 m}c_{1\ol
m}\rangle=\langle c_{3 m}c_{2\ol m}\rangle=\langle c_{1 m}c_{3\ol
m}\rangle=\langle c_{2\ol m}c_{1 m}\rangle=\langle c_{3\ol m}c_{2
m}\rangle=\langle c_{1\ol m}c_{3 m}\rangle=:D_m$, where $D_m$ is
real. For $0<m\leq\Omega'$ we have
\begin{eqnarray*}&\langle c_{2\ol m}c_{1
m}\rangle&=K-K^2\left(\langle c_{2\ol m}^\dagger c_{3
m}^\dagger\rangle+\langle c_{3\ol m}^\dagger c_{1
m}^\dagger\rangle+\langle c_{1\ol m}^\dagger c_{2
m}^\dagger\rangle-\langle c_{3\ol m}^\dagger c_{3
m}^\dagger\rangle\right),\\
&\langle c_{1\ol m} c_{1 m}\rangle&=-K^2\left(\langle c_{2\ol
m}^\dagger c_{2 m}^\dagger \rangle +\langle c_{3\ol m}^\dagger c_{3
m}^\dagger\rangle-\langle c_{3\ol m}^\dagger c_{2
m}^\dagger\rangle-\langle c_{2 m}^\dagger c_{3\ol
m}^\dagger\rangle\right),\\&& =-K^2\left(\langle c_{2\ol m}^\dagger
c_{2 m}^\dagger \rangle +\langle c_{3\ol m}^\dagger c_{3
m}^\dagger\rangle\right) ,
\end{eqnarray*}
 which imply
$$D_m=K-3K^2D_m+K^2P_m,\quad P_m=2K^2P_m,
$$ where $P_m:=\langle c_{1 m}c_{1\ol m}\rangle=\langle c_{2 m}c_{2\ol m}\rangle=\langle c_{3 m}c_{3\ol
m}\rangle,$ is also real. The procedure for $\Omega'<m\leq\Omega$,
is analogous. Finally, we find $P_m=0 $ and
$$D_m={K\over1+3K^2},\quad 0<m\leq\Omega';\quad D_m=
{\tilde K\over1+3\tilde K^2},\quad \Omega'<m\leq\Omega.
$$
 We are now able to compute the energy expectation value $${{\cal E}\over
 G}=
\sum_{j=1}^3{\langle\Phi| A^\dagger_{j} A_{j}|\Phi\rangle\over
\langle\Phi|\Phi\rangle} 
.$$
\bigskip
\subsection{ Computation of the energy expectation value}
\begin{figure}[ht]
\centering
\includegraphics[width=0.6\textwidth, height=0.5\textwidth]{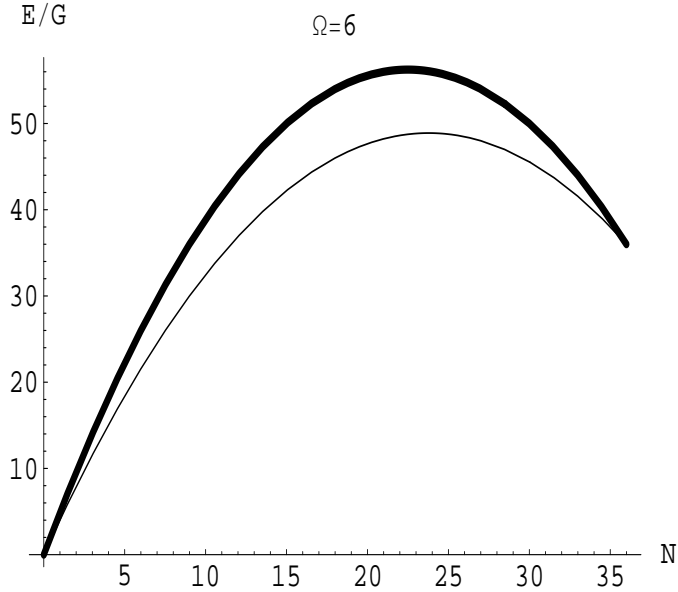}
\caption{Groundstate energy of the color symmetrical sector versus
the quark number, for $\Omega=6$. Thick line: exact result according
to (\ref{8}); thin line: color symmetrical BCS estimate according to
(\ref{5}). Since $G<0$, the maximum of $E/G$ corresponds to the
minimum of $E$.} \label{fig1}
\end{figure}
\begin{figure}[ht]
\centering
\includegraphics[width=0.6\textwidth, height=0.5\textwidth]{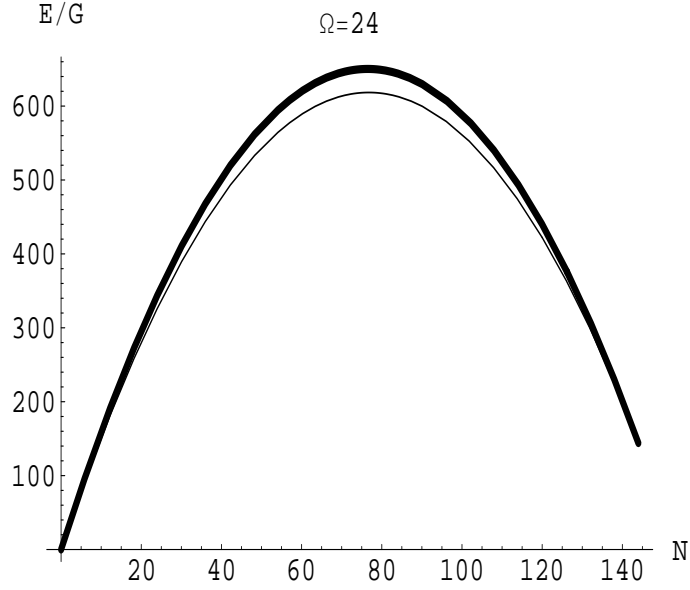}
\caption{Groundstate energy of the color symmetrical sector versus
the quark number, for $\Omega=24$. Thick line: exact result
according to (\ref{8}); thin line: color symmetrical BCS estimate
according to (\ref{5}).} \label{fig2}
\end{figure}
Let $p=\Omega'/\Omega,~q=1-p=(\Omega-\Omega')/\Omega.$ Define
$\theta,~\tilde\theta$ such that
$\sqrt{3}K/\sqrt{1+3K^2}=\sin{\theta},~1/\sqrt{1+3K^2}=\cos{\theta},
~\sqrt{3}\tilde
 K/\sqrt{1+3\tilde  K^2}=\sin{\tilde\theta},~1/\sqrt{1+3\tilde  K^2}=\cos{\tilde
\theta}.$
 Then, the color symmetry constraint reads
 $$p\cos2\theta-q\cos2\tilde\theta=p-q.
 $$ The main contribution to the energy expectation value comes from
 the square of the following expectation values, which involve
 contractions of the form $\langle cc\rangle$,
 $$\langle A_1\rangle=\langle A_2\rangle=\langle
 A_3\rangle={\Omega\over\sqrt{3}}(p\sin2\theta+q\sin2\tilde\theta).
 $$ The corresponding constrained extremum occurs for
 $$\cos2\theta=-\cos2\tilde\theta=p-q,~\sin2\theta=\sin2\tilde\theta=\sqrt{1-(p-q)^2}, $$
 so that, in the leading order, ${\cal E}/G\approx\Omega^2(1-(p-q)^2)$. In
 terms of the variables $\theta,~\tilde\theta,$ the number of quarks
 reads $N=6\Omega[p(1-\cos2\theta)/3+q(1-(1-\cos2\tilde\theta)/3)].$
 At the extremum, $N=6\Omega q$, and ${\cal E}/G\approx\Omega^2[1-(1-N/(3\Omega))^2].
 $ To complete the calculation of ${\cal E}/G$ we must add the small
 corrections coming from the neglected contractions of the form $\langle c^\dagger
 c\rangle$.

In terms of $\theta,~\tilde\theta$, we have
$$\langle c_{jm}^\dagger c_{jm}\rangle=\langle c_{j\ol
m}^\dagger c_{j\ol m}\rangle=
{1\over3}(1-\cos2\theta),\quad 0<m\leq\Omega',$$
$$
\quad \langle c_{jm}^\dagger c_{jm}\rangle=\langle c_{j\ol
m}^\dagger c_{j\ol m}\rangle=1-
{1\over3}(1-\cos2\tilde\theta),\quad \Omega'<m\leq\Omega.$$
 At the
extremum,
$$\langle c_{jm}^\dagger c_{jm}\rangle=\langle c_{j\ol
m}^\dagger c_{j\ol m}\rangle= {2\over3}~q ,\quad 0<m\leq\Omega',$$
$$
\quad \langle c_{jm}^\dagger c_{jm}\rangle=\langle c_{j\ol
m}^\dagger c_{j\ol m}\rangle={1\over3}(1 +2q),\quad
\Omega'<m\leq\Omega.$$ The contribution, mentioned above, of the
neglected contractions of the form $\langle c^\dagger c\rangle$, to
${\cal E}/G$, is
$$6\Omega\left[p\left({2\over3}~q\right)^2+\left({1\over3}(1
+2q)\right)^2\right]={2\Omega
q\over3}(1+8q)={N\over9}\left(1+{4N\over3\Omega}\right).
$$ Finally, the groundstate energy of the color symmetrical
super-conducting phase reads,
\begin{equation}\label{5}{{\cal E}\over
G}={N\over9}\left(6\Omega-{N}+1+{4N\over3\Omega}\right),\quad 0\leq
N\leq6\Omega. \end{equation} Although eq. (\ref{5}) is close to eq.
(29) of ref. \cite{bohr}, showing the same qualitative behavior,
this agreement is, to some extent, accidental.

Using the Schwinger type representation of $su(4)$, formulated in
terms of appropriate boson operators, which was developed by
Yamamura {\it at al.} \cite{yamamura}, the color symmetrical sector
of the Bonn model has been characterized in \cite{c.providencia}.
There, the exact groundstate energy of the color symmetrical sector
was found to read
\begin{equation}\label{8}{{\cal E}\over
G}={N\over3}\left(2\Omega+3-{N\over3}\right),\quad 0\leq
N\leq6\Omega. \end{equation} It is interesting to compare eq.
(\ref{5}) with eq. (\ref{8}), because the comparison shows that the
approximate result of (\ref{5}) is in reasonable agreement with the
exact result of (\ref{8}).

\section{Conclusions}
We have constructed a BCS-type formalism, based on a conveniently
generalized Bogoliubov transformation, which is appropriate to
describe color symmetrical superconducting states of quark matter.
It is interesting to compare eq. (\ref{5}), showing the BCS estimate
of the groundstate energy of the color symmetric sector, and eq.
(\ref{8}), showing the exact groundstate energy of the same sector.
This is done, in fig. \ref{fig1}, for $\Omega=6$ and in fig.
\ref{fig2}, for $\Omega=24$. It is clearly seen that the color
symmetric BCS result becomes closer to the exact one while $\Omega$
increases.

It should be emphasized that the present approach automatically
ensures vanishing net color charge, in the average.

A word is in order concerning the mechanism of color-flavor-locking
(CFL) which was  introduced in \cite{alford} and does lead to color
symmetric superconductivity, in the average. However, CFL is only
meaningful if the three flavors $u,d,s$ are equally relevant. The
new form of color symmetrical superconductivity which is proposed
here is, of course, distinct from CFL, and is also meaningful when
only the flavors $u,d$ are relevant. The CFL property considered in
ref.  \cite{alford} is destroyed by an arbitrary infinitesimal color
rotation, unless eq. (\ref{cneutral}) is implemented.


\section*{Acknowledgements}
One of the authors (JP) wishes to acknowledge most valuable
discussions with Prof. Mastoshi Yamamura. The present research was
partially supported by Project POCI/FP/81923/2007.

\end{document}